\documentstyle[12pt]{article}
\begin{document}
\begin{center}
{\bf Influence of Non-Resonant Effects on the Dynamics \\
of Quantum Logic Gates at Room Temperature}\\ \ \\
G.P. Berman$^1$, A.R. Bishop$^1$, G.D. Doolen$^1$, G.V. L\'opez$^2$, and V.I. Tsifrinovich$^3$\\ \ \\
\end{center}
$^1$Theoretical Division and CNLS, Los Alamos National Laboratory, \\
Los Alamos, New Mexico 87545\\
$^2$ Departamento de F\'isica, Universidad de Guadalajara,\\
Corregidora 500, S.R. 44420, Guadalajara, Jalisco, M\'exico\\
$^3$Department of Applied Mathematics and Physics, Polytechnic University,\\
Sixth Metrotech Center,
Brooklyn NY 11201\\ \ \\
\begin{center}
{\bf ABSTRACT}
\end{center}
We study numerically the influence of non-resonant effects on the dynamics of a single $\pi$-pulse quantum CONTROL-NOT (CN) gate in a macroscopic ensemble of four-spin molecules at room temperature. The four nuclear spins in each molecule represent a four-qubit register. The qubits are ``labeled'' by the characteristic frequencies, $\omega_k$, ($k=0$ to $3$) 
due to the Zeeman interaction of the nuclear spins with the magnetic field. The qubits interact with each other through an Ising interaction of strength  $J$. The paper examines the feasibility of implementing a single-pulse quantum CN gate in an ensemble of quantum molecules at room temperature. We determine a parameter region,  $\omega_k$ and $J$, in which a single-pulse quantum CN gate can be implemented at room temperature. We also show that there exist characteristic critical values of parameters, $\Delta\omega_{cr}\equiv|\omega_{k^\prime}-\omega_k|_{cr}$ and $J_{cr}$, such that for $J<J_{cr}$ and $\Delta\omega_k\equiv|\omega_{k^\prime}-\omega_k|<\Delta\omega_{cr}$, non-resonant effects are sufficient to destroy the dynamics required for quantum logic operations. 
\newpage
\quad\\
{\bf I. Introduction}\\ \ \\
Recently, the problems related to quantum computation have attracted much attention. (See, for example, the reviews and books on this subject, \cite{lloyd1}-\cite{hughes}, and references therein.) Instead of bits with  values ``0'' and ``1'' (the states of a transistor) which are used in a classical computer, a quantum computer uses quantum bits -- qubits. Usually, a qubit is implemented in a quantum two-level system such as a nuclear or electron spin, an electron in an atom (in an ion, or in a quantum dot), or a photon in a cavity. The wave function of an individual qubit can be prepared in the ground state, $|0\rangle$, or in the excited state, $|1\rangle$, or in any superpositional state: $\Psi=c_0|0\rangle + c_1|1\rangle$. (The only restriction is the normalization condition: $|c_0|^2+|c_1|^2=1$.) Superpositional states provide significant advantages for quantum computation.

Consider, for example, a molecule (a ``register'') containing four 
nuclear spins (qubits) which we shall label by ``0'', ``1'', ``2'', and ``3'', from the right to the left. This quantum register can be prepared initially in the state,
$$
\Psi={{1}\over{\sqrt{2}}}(|0_3\rangle+|1_3\rangle)\otimes{{1}\over{\sqrt{2}}}(|0_2\rangle+|1_2\rangle)\otimes{{1}\over{\sqrt{2}}}(|0_1\rangle+|1_1\rangle)\otimes{{1}\over{\sqrt{2}}}(|0_0\rangle+|1_0\rangle)\equiv
$$
$$
{{1}\over{4}}(|0_30_20_10_0\rangle+|0_30_20_11_0\rangle+...+|1_31_21_11_0\rangle).\eqno(1.1)
$$
In decimal notation, the state (1.1) can be written as,
$$
\Psi={{1}\over{4}}(|0\rangle+|1\rangle+...+|15\rangle).\eqno(1.2)
$$
It follows from (1.1) and (1.2) that already at the initial step of a quantum computation, one can load a four-qubit register with {\it all} numbers, from ``0'' to ``15''. This is not possible in a classical computer. 

After loading, one can perform quantum calculations using unitary transformations of the wave function (1.1). These calculations could utilize, for example, recently discovered Shor \cite{shor} or Grover \cite{grov2} algorithm. This step of quantum computation usually requires error correction codes \cite{knill,steane2}. The final states of qubits are read using conventional techniques \cite{chuang98}. (See also strategies for mesoscopic measuring devices \cite{gur1}-\cite{stod}.) All these operations must be completed at times smaller than the time of decoherence. This requirement places significant limitations on the class of materials which can be used for quantum computation \cite{vinc1}. The first quantum computation (implementation of Grover's algorithm) has been demonstrated using a nuclear magnetic resonance (NMR) techniques at room temperature \cite{chuang98}. In this case, the qubits were nuclear spins 
in a molecule. 

The states which are used for quantum computation are generally non-stationary states because they are not the eigenstates of the Hamiltonian describing the quantum computer. Understanding of the dynamics of these states is important.  Resolving the dynamical issues of quantum computation is required in order to build a working quantum computer. One of these issues is the importance of non-resonant interactions.
The dynamics of quantum computation involves both resonant and non-resonant interactions. Under some conditions, non-resonant effects can be ignored. In our previous paper \cite{berm2}, we studied the dynamics of the quantum CN gate in a parameter regime in which the non-resonant effects can be neglected. 
But in quantum computation non-resonant effects can accumulate. Even if the interaction between qubits is small, non-resonant effects can inhibit the desired effects and create significant errors. The design of a quantum computer thus requires knowing those regions of parameter space in which non-resonant effects are negligible. 

In this paper, we study numerically the influence of non-resonant effects on the dynamics of a single-pulse quantum CN gate in an ensemble of four-spin molecules at room temperature. We calculate the region of parameters in which  a single-pulse quantum CN gate can be implemented at room temperature. 

The quantum CN gate is of central importance in quantum
computation because any quantum logic gate can be decomposed into a
sequence of one-qubit rotations and two-qubit quantum CN gates. Recent
implementation of the quantum CN gate in a liquid state NMR involved a special
sequence of resonant electromagnetic pulses. For example, in the
``proton-carbon'' nuclear spin system, the quantum CN gate was implemented using two radio-frequency pulses \cite{chuang}. The first pulse induced a $\pi$/2 rotation of the target carbon spin around the one axis of
the rotating reference frame. The second pulse induced a similar
rotation around the other axis. The delay time between two pulses was
$\pi/2J$, where $J$ is the interaction constant. The action of the
second pulse depends on the state of the control proton spin. In
particular, the second pulse cancels the action of the first pulse if the proton spin is in the ground state. In this experiment, the resonant frequencies were:
500 MHz for the proton spin, and 125 MHz for the carbon spin, $J/2\pi$ = 108
Hz. In the case of a single-pulse quantum CN gate, the Rabi frequency, $\Omega$, must be less than the interaction constant, $J$, to provide a highly selective excitation of spin levels. For a two-pulse quantum CN gate, one should control the amplitude, duration and phases of two pulses, and also the delay time between the pulses. A single-pulse quantum CN gate is much simpler for implementation.  But it takes more time than a two-pulse quantum CN gate because for a single-pulse quantum CN gate $\Omega\ll J$, and the pulse duration is: $\tau=\pi/\Omega$. We show that a single-pulse quantum CN gate could be implemented in liquid state NMR and in solid-state systems in which the selective excitations can be realized \cite{lloyd1,lloyd2,kane,cl}. 
Below we assume that the molecules are in a constant magnetic field. The four spins in a single molecule interact with each other through Ising interactions. Their nuclear spins interact also with electromagnetic pulses. In Section II, we present the Hamiltonian of this system, introduce a quantum CN gate for our system, and discuss the initial conditions for the density matrix. In Section III, we present the results of numerical calculations. We investigate the influence of non-resonant effects on the dynamics of a quantum CN gate and calculate the region of parameters in which  a single-pulse CN gate can be implemented. In the Conclusion, we summarize our results. \\ \ \\
{\bf II. Equations of Motion}\\ \ \\
Consider an ensemble of non-interacting four-spin molecules which are placed in the magnetic field,
$$
{\bf B}=(B_\perp\cos\omega t,-B_\perp\sin\omega t,B_\parallel).\eqno(2.1)
$$
In (2.1), $B_\parallel$ is the $z$-component of the magnetic field, and $B_\perp$ is the amplitude of the circularly polarized magnetic field which rotates in the $(x,y)$ plane with the frequency, $\omega$. It is convenient to represent the four spins in a single molecule by four spin indices: $i_3,j_2,k_1,l_0$, where the numbers $3$, $2$, $1$, and $0$, indicate positions of nuclear spins ($I=1/2$) in a four-qubit ``register''. The indices,
$i_3$, $j_2$, $k_1$, and $l_0$ take the values, ``0'' (orientation of the corresponding spin in the positive $z$-direction), or ``1''  (orientation of the corresponding spin in the negative $z$-direction). All spins interact with each 
other through the Ising interaction with the constants of interaction, $J_{\alpha,\beta}$ ($\alpha,\beta=0,1,2,3$). The Hamiltonian of the system is,
$$
{\cal H}=H+V,\eqno(2.2)
$$
$$
H=-\hbar\sum_{\alpha=0}^3[\omega_\alpha I^z_\alpha +2\sum_{\beta>\alpha}J_{\alpha,\beta}
I^z_\alpha I^z_\beta],\eqno(2.3)
$$
$$
V=-{{\hbar}\over{2}}\sum_{\alpha=0}^3\Omega_\alpha\Bigg(e^{i\omega t}I^+_\alpha+e^{-i\omega t}I^-_\alpha\Bigg).
$$
In (2.2), (2.3), $\omega_\alpha=\gamma_\alpha B_\parallel$, is the frequency of the spin precession due to the Zeeman interaction; $\Omega_\alpha=\gamma_\alpha B_\perp$, is the Rabi frequency; $I_\alpha^{+,-}=I_\alpha^x\pm iI_\alpha^y$; $I_\alpha^{x,y,z}=(1/2)\sigma^{x,y,z}_\alpha$ (where $\gamma_\alpha$ is the gyromagnetic ratio, and $\sigma_\alpha^{x,y,z}$ is the Pauli matrix, $\alpha=0,1,2,3$). The Hamiltonian ${\cal H}$ is used to solve the equation for the density matrix: $i\hbar\dot\rho=[{\cal H},\rho]$.\\ \ \\
{\bf Rotating system of coordinates}\\ 
The Hamiltonian ${\cal H}$ is time-dependent. This time dependence can be eliminated by transforming to a system of coordinates which rotates with the frequency of the external magnetic field, $\omega$. This transformation can be made using the unitary operator, $U_\omega=\exp[i\omega t(\sum_{\alpha=0}^3I^z_\alpha)]$. In this rotating system of coordinates, we have the time-independent Hamiltonian,
$$ 
{\cal H}^\prime=U_\omega^\dagger{\cal H}U_\omega=H^\prime+V^\prime,\eqno(2.4)
$$
$$
H^\prime=-\hbar\sum_{\alpha=0}^3[(\omega_\alpha-\omega)I^z_\alpha+2\sum_{\beta>
\alpha}J_{\alpha,\beta}I^z_\alpha I^z_\beta],
$$
$$
V^\prime=-\hbar\sum_{\alpha=0}^3\Omega_\alpha I^x_\alpha.
$$
The equation for the new density matrix, $\rho^\prime=U_\omega^\dagger\rho U_\omega$, is: $i\hbar\dot\rho^\prime=[{\cal H}^\prime,\rho^\prime]$. To calculate the density matrix, $\rho^\prime$, we use the complete set of the basis states, $|i_3j_2k_1l_0\rangle$,
$$
|0000\rangle\equiv |0_30_20_10_0\rangle=|0\rangle,\quad 
|0001\rangle\equiv |0_30_20_11_0\rangle=|1\rangle, ..., 
|1111\rangle\equiv |1_31_21_11_0\rangle=|15\rangle .\eqno(2.5)
$$
The sixteen states, $|i_3j_2n_1m_0\rangle$, form a complete set of eigenstates of the Hamiltonian $H^\prime$ in (2.4): $H^\prime|k\rangle=E^\prime_k|k\rangle$, ($k=0,...,15$). For example, in the rotating system of coordinates, the energies of the ground state, $|0\rangle$, and the first excited state, $|1\rangle$, are,
$$
E_0^\prime=-{{\hbar}\over{2}}\sum_{\alpha=0}^3[(\omega_\alpha-\omega)+\sum_{\beta>\alpha}J_{\alpha,\beta}].\eqno(2.6)
$$
$$
E_1^\prime=-{{\hbar}\over{2}}(-\omega_0+\omega_1+\omega_2+\omega_3-2\omega-
J_{0,1}-J_{0,2}-J_{0,3}+J_{1,2}+J_{1,3}+J_{2,3}).
$$
We assume that: $\omega_0<\omega_1<\omega_2<\omega_3$, and $J\ll \omega_{\alpha+1}-\omega_\alpha$, ($\alpha=0,1,2$).\\ \ \\
{\bf Excluding fast dynamics}\\ 
Despite the fact that the equation for the density matrix, $\rho^\prime$, has time-independent coefficients,
this equation is still not convenient for numerical calculations. This equation includes both fast and slow dynamics. The fast dynamics is associated with the presence of high frequency terms $\sim(\omega_\alpha-\omega$). The slow dynamics is associated with terms proportional to the Rabi frequencies, $\sim\Omega_\alpha\ll\omega_\alpha$. For numerical calculation of the 
density matrix, it is convenient to exclude high-frequency oscillations for each qubit. To do this, we use the unitary transformation: $U=\exp(-iH^\prime t/\hbar)$. Then, we have the following equation for the new density matrix,
$$
i\hbar \dot\rho^{\prime\prime}=[V^{\prime\prime},\rho^{\prime\prime}],\quad (\rho^{\prime\prime}=U^\dagger\rho^\prime U,\quad V^{\prime\prime}=U^\dagger V^\prime U).\eqno(2.7)
$$
Note that $V^{\prime\prime}$ is a time-dependent operator.
\\ \ \\
{\bf Initial conditions}\\
The thermal equilibrium initial conditions for the density matrix (2.7) are,
$$
\rho^{\prime\prime}_{kk}(0)=\rho_{kk}(0)={{e^{-E_k/k_BT}}\over{\sum_{k=0}^{15}e^{-E_k/k_BT}}},\quad \rho^{\prime\prime}_{ik}(0)=0, \quad (i,k=0-15,\quad i\not= k),\eqno(2.8)
$$
where $E_k$ are the eigenvalues of the Hamiltonian $H$ in (2.3) ($E_k=E_k^\prime(\omega=0)$).
The expression (2.8) can be simplified using the condition, $E_k/k_BT\ll 1$, 
which is usually satisfied for an ensemble of nuclear spins at room temperature.
We assume also that the difference between the single-spin transition frequencies is small compared with the average frequency, $\sum^3_{\alpha=0}\omega_\alpha/4$. Besides, following the idea suggested in \cite{gers}, we assume that one can use a sequence of electromagnetic pulses which allows one to prepare the initial density matrix (2.8) in the state,
$$
\rho=E/16+\rho_\Delta,\eqno(2.9)
$$
$$
\rho_\Delta={{\hbar\sum_{k=0}^3\omega_k}\over{32k_BT}}\Bigg[|0><0|+{{1}\over{2}}\Bigg(-|4><4|+|5><5|+|6><6|+|7><7|+
$$
$$
|8><8|-|9><9|-|10><10|-|11><11|\Bigg)-|12><12|\Bigg].
$$
In (2.9), $E$ is the unit matrix, $\rho_\Delta$ is the deviation matrix, and $Tr(\rho_\Delta)=0$. The idea suggested in \cite{gers}, was to use the first four basic states, $|00kl\rangle $ ($k,l=0,1$), as the ``active states'' for quantum logic operations. This density matrix in the form (2.9) is chosen because, the dynamics of this ensemble of four-spin molecules is formally identical (up to some extent) to the dynamics of a pure quantum system of two interacting spins. Below, we shall assume (2.9) as the initial state.\\ \ \\
{\bf Quantum CONTROL-NOT (CN) gate}\\
Assume that the frequency of the external magnetic field, $\omega$, is resonant with the frequency of the transition, $|0\rangle\leftrightarrow|1\rangle$, and that this transition frequency is different from all other single-spin transition frequencies. In this case, by applying a single $\pi$-pulse with frequency $\omega$, we shall implement the modified quantum CN gate in an ensemble of four-spin molecules,
$$
\hat{CN}=i|0000\rangle\langle 0001|+i|0001\rangle\langle 0000|+\sum_{p,q,r,s=0,1}|pqrs\rangle\langle pqrs|\equiv\eqno(2.10)
$$
$$
i|0\rangle\langle 1|+i|1\rangle\langle 0|+\sum_{n=2}^{15}|n\rangle\langle n|,\quad |pqrs\rangle\not=|0000\rangle,|0001\rangle.
$$
The operation (2.10) changes the state of the right-most spin (the ``target'' spin: ``0''$\leftrightarrow$ ``1'', only if its neighbor (the ``control'' spin) is in the state ``0''. If we assume that the $\hat{CN}$ gate in an ensemble of four-spin molecules works analogously to the   $\hat{CN}$ gate in a pure quantum two-spin system, then some conditions on the density matrix must be satisfied. We introduce the initial wave function for the corresponding pure quantum two-spin system,
$$
|\Psi^{\prime\prime}(0)\rangle=c^{\prime\prime}_0(0)|0\rangle+c^{\prime\prime}_1(0)|1\rangle+
c^{\prime\prime}_2(0)|2\rangle+c^{\prime\prime}_3(0)|3\rangle.\eqno(2.11)
$$
It follows from (2.10) and (2.11), that after the action of a resonant $\pi$-pulse, one has the following wave function,
$$
|\Psi^{\prime\prime}(\tau)\rangle=c^{\prime\prime}_0(\tau)|0\rangle+c^{\prime\prime}_1(\tau)|1\rangle+
c^{\prime\prime}_2(\tau)|2\rangle+c^{\prime\prime}_3(\tau)|3\rangle=\eqno(2.12)
$$
$$
\hat{CN}|\Psi^{\prime\prime}(0)\rangle=ic^{\prime\prime}_0(0)|1\rangle+ic^{\prime\prime}_1(0)|0\rangle+
c^{\prime\prime}_2(0)|2\rangle+c^{\prime\prime}_3(0)|3\rangle,
$$
where the duration of a $\pi$-pulse is approximately,
$$
\tau={{\pi}\over{\Omega_0}}.\eqno(2.13)
$$
Note, that (2.13) defines the time of the $\pi$-pulse only for a one-qubit rotation. For the complicated system under consideration, (2.13) gives only an approximate value for the time of a $\pi$-pulse.

It follows from (2.12) that the following conditions should be satisfied,
$$
c^{\prime\prime}_0(\tau)=ic_1^{\prime\prime}(0),\quad 
c^{\prime\prime}_1(\tau)=ic_0^{\prime\prime}(0),\quad
c^{\prime\prime}_2(\tau)=c_2^{\prime\prime}(0),\quad 
c^{\prime\prime}_3(\tau)=c_3^{\prime\prime}(0).\eqno(2.14)
$$
Now, we express the conditions (2.14) in terms of the density matrix. We introduce the density matrix for a pure two-spin system,
$$
r^{\prime\prime}_{nk}(0)\equiv c^{\prime\prime}_n(0){c_k^*}^{\prime\prime}(0),\quad (n,k=0-3).\eqno(2.15)
$$
>From (2.14) and (2.15) we derive the following relation between the density matrix elements of the matrices, $r^{\prime\prime}(\tau)$ and $r^{\prime\prime}(0)$,
$$
r^{\prime\prime}(\tau)=
\left(\matrix{
r^{\prime\prime}_{00}(\tau) & r^{\prime\prime}_{01}(\tau) & r^{\prime\prime}_{02}(\tau) & r^{\prime\prime}_{03}(\tau) \cr 
r^{\prime\prime}_{10}(\tau) & r^{\prime\prime}_{11}(\tau) & r^{\prime\prime}_{12}(\tau) & r^{\prime\prime}_{13}(\tau) \cr 
r^{\prime\prime}_{20}(\tau) & r^{\prime\prime}_{21}(\tau) & r^{\prime\prime}_{22}(\tau) & r^{\prime\prime}_{23}(\tau) \cr 
r^{\prime\prime}_{30}(\tau) & r^{\prime\prime}_{31}(\tau) & r^{\prime\prime}_{32}(\tau) & r^{\prime\prime}_{33}(\tau) \cr 
}\right)=
\left(\matrix{
r^{\prime\prime}_{11}(0) & r^{\prime\prime}_{10}(0) & ir^{\prime\prime}_{12}(0) & ir^{\prime\prime}_{13}(0) \cr 
r^{\prime\prime}_{01}(0) & r^{\prime\prime}_{00}(0) & ir^{\prime\prime}_{02}(0) & ir^{\prime\prime}_{03}(0) \cr 
-ir^{\prime\prime}_{21}(0) & -ir^{\prime\prime}_{20}(0) & r^{\prime\prime}_{22}(0) & r^{\prime\prime}_{23}(0) \cr 
-ir^{\prime\prime}_{31}(0) & -ir^{\prime\prime}_{30}(0) & r^{\prime\prime}_{32}(0) & r^{\prime\prime}_{33}(0) \cr 
}\right).\eqno(2.16)
$$
All other matrix elements of the initial density matrix, $\rho_\Delta^{\prime\prime}(0)$, in (2.9) are assumed to not change significantly in the process of the quantum CN operation. The relations (2.16) will be verified numerically in Section III for an ensemble of four-spin molecules.

In our numerical experiment, we calculated the dynamics of the density matrix,
$$
\rho_\Delta(t)={{\hbar\sum_{k=0}^3\omega_k}\over{32k_BT}}\sum_{n,k=0}^{15}r_{n,k}^{\prime\prime}(t)|n\rangle\langle k|,\eqno(2.17)
$$
with the initial condition,
$$
r^{\prime\prime}(0)=\sum^3_{n,k=0}r_{nk}(0)|n><k|+{{1}\over{2}}\Bigg(-|4><4|+|5><5|+|6><6|+\eqno(2.18)
$$
$$
|7><7|+|8><8|-|9><9|-|10><10|-|11><11|\Bigg)-|12><12|.
$$
\\ \ \\
{\bf III. Results of Numerical Calculations}\\ \ \\
 In this section, we study numerically the dynamics of the quantum CN gate, and the influence of non-resonant effects on the dynamics of the quantum CN gate. We calculate the slow dynamics for the density matrix, $r^{\prime\prime}$.
 Below, the upper index, $^{\prime\prime}$, is omitted. The following parameters were chosen,
$$
\omega_k=\omega_0+100\times k,
\quad J_{\alpha,\beta}=J=10,\quad (k,\alpha,\beta=0,1,2,3)\eqno(3.1)
$$
$$
\omega=(E_1-E_0)/\hbar=\omega_0+3J,\quad \Omega_k=\Omega=0.1.
$$
(The characteristic dimensional parameters can be obtained by multiplying the values in (3.1) by $2\pi\times 10^6s^{-1}$.) 
The condition $\omega=(E_1-E_0)/\hbar=\omega_0+3J$ (or $(E_1^\prime-E_0^\prime )/\hbar=0$) corresponds to the resonant transition $|0\rangle\leftrightarrow |1\rangle$. As was mentioned above, the frequency of this transition differs from the frequencies of all other single-spin transitions. The density matrix in the form, 
$$
r(0)=\left(\matrix{
1&0&0&0\cr
0&0&0&0\cr
0&0&0&0\cr
0&0&0&0\cr
}\right),\eqno(3.2)
$$
corresponds to the initial state, $\Psi(0)=|00\rangle$, of a pure quantum system. In (3.2), we presented only the part of the density matrix (we shall call it below, ``reduced density matrix'' (RDM)) with the components, $r_{ij}$ ($i,j=0-3$). All other components, here and below, are given in (2.18).\\ \ \\
{\bf Dynamics of superpositional state}\\ 
The quantum CN gate (2.10) should work for {\it any} initial state, $\Psi(0)$. In this sub-section, we consider the dynamics of the quantum CN gate (2.10) for a superpositional initial state. We choose here the initial RDM:
$$
r(0)=\left(\matrix{
0.3&0.2449&0.3162&0.2236\cr
0.2449&0.2&0.2582&0.1826\cr
0.3162&0.2582&0.333&0.2357\cr
0.2236&0.1826&0.2357&0.1666\cr
}\right).\eqno(3.3)
$$
The RDM (3.3) corresponds to the following initial wave function of a pure two-spin quantum system,
$$
\Psi(0)=\sqrt{0.3}|00\rangle+\sqrt{0.2}|01\rangle+{{1}\over{\sqrt{3}}}|10\rangle+{{1}\over{\sqrt{6}}}|11\rangle.\eqno(3.4)
$$
Now, we apply a $\pi$-pulse with frequency, $\omega=\omega_0+3J$. It follows from (2.16) that, according to the evolution of a pure quantum system, one should expect the following RDM at the end of the $\pi$-pulse, 
$$
r(t=\pi/\Omega_0)=\left(\matrix{
0.2&0.2449&i0.25819&i0.1826\cr
0.2449&0.3&i0.3162&i0.2236\cr
-i0.2582&-i0.3162&0.333&0.2357\cr
-i0.1826&-i0.2236&0.2357&0.1666\cr
}\right).\eqno(3.5)
$$
The RDM (3.5) corresponds to the following wave function of a pure quantum system:\\ $\Psi(t=\pi/\Omega_0)=i\sqrt{0.2}|00\rangle+i\sqrt{0.3}|01\rangle+(1/\sqrt{3})|10\rangle+(1/\sqrt{6})|11\rangle$. 

The results of numerical calculations of the dynamics of the initial RDM (3.3) for parameters given in (3.1), are shown in Figs 1-3. As one can see from Figs 1-3, the values of the matrix elements of the RDM at the end of a $\pi$-pulse approximately coincide with the expected solution (3.5). The deviation is about
0.5\%, and is caused by non-resonant effects.\\ \ \\
{\bf Influence of non-resonant effects}\\
As shown above, for some regions of parameters, the influence of non-resonant effects on the dynamics of a quantum CN gate leads only to small effects (about 0.5\%). In this case, one can realize (at least for a finite time) quantum logic operations in an ensemble of four-spin molecules described by the Hamiltonian (2.3). The reason the non-resonant effects are small in the numerical calculations described above, is that the parameters (3.1) which we used, allowed us: (a) to separate significantly the frequencies, $\omega_k$, for different qubits (to ``label'' the qubits); and (b) to choose the interaction constant, $J$, between the qubits significantly large: $J\gg \Omega$. Indeed, if $J\ll \Omega$, the single-spin transitions become degenerate. For example, the frequency of the transition $|0000\rangle\leftrightarrow|0001\rangle$ is $\omega_0+3J$; and the frequency of the three transitions $|0010\rangle\leftrightarrow|0011\rangle$, $|0100\rangle\leftrightarrow|0101\rangle$, !
!
and $|1000\rangle\leftrightarrow|1001\rangle$ is $\omega_0+J$. If the value of $J$ becomes small, all these transitions become degenerate, and they all  contribute into the dynamics of the system. This degeneracy destroys quantum logic operations. Below, in this sub-section, we shall investigate numerically the region of parameters $\Delta\omega_k=|\omega_{k^\prime}-\omega_{k}|$ and $J$, which separates significant from insignificant influence of non-resonant effects on the dynamics of the quantum CN gate. 

The matrix elements are complex functions which are time-dependent. They include both time-dependent amplitudes and phases: $r_{nk}(t)=|r_{nk}(t)|\exp(i\phi_{nk}(t))$, where the time of a $\pi$-pulse, $\tau$, is defined in (2.13).
  To study the deviation of the matrix elements from their desired (expected) values (2.16),  due to non-resonant effects, we introduce the deviations of the amplitudes, $\Delta_{ij}$, and the phases, $\delta_{ij}$,
$$ 
 \Delta_{00}(\tau)=||r_{00}(\tau)|-|r_{11}(0)||,\quad  
\Delta_{01}(\tau)=||r_{01}(\tau)|-|r_{10}(0)||,\quad 
 \Delta_{02}(\tau)=||r_{02}(\tau)|-|r_{12}(0)||, 
$$
$$
\Delta_{03}(\tau)=||r_{03}(\tau)|-|r_{13}(0)||,\quad 
 \Delta_{11}(\tau)=||r_{11}(\tau)|-|r_{00}(0)||,\quad  
\Delta_{12}(\tau)=||r_{12}(\tau)|-|r_{02}(0)||, 
 $$
$$
\Delta_{13}(\tau)=||r_{13}(\tau)|-|r_{03}(0)||,\quad  
\Delta_{22}(\tau)=||r_{22}(\tau)|-|r_{22}(0)||,\quad 
 \Delta_{23}(\tau)=||r_{23}(\tau)|-|r_{23}(0)||, 
$$
$$
\Delta_{33}(\tau)=||r_{33}(\tau)|-|r_{33}(0)||,\eqno(3.6)
$$
and $\Delta_{ii}(\tau) =||r_{ii}(\tau)|-|r_{ii}(0)||$, for $15\ge i\ge 4$. For the deviation of phases we have, according to (2.16),
$$
\delta_{00}(\tau)=|\phi_{00}(\tau)-\phi_{11}(0)|,\quad
\delta_{01}(\tau)=|\phi_{01}(\tau)-\phi_{10}(0)|,\quad
\delta_{02}(\tau)=|\phi_{02}(\tau)-\phi_{12}(0)-\pi/2|,\quad
$$
$$
\delta_{03}(\tau)=|\phi_{03}(\tau)-\phi_{13}(0)-\pi/2|,\quad
\delta_{11}(\tau)=|\phi_{11}(\tau)-\phi_{00}(0)|,\quad
$$
$$
\delta_{12}(\tau)=|\phi_{12}(\tau)-\phi_{02}(0)-\pi/2|,\quad
\delta_{13}(\tau)=|\phi_{13}(\tau)-\phi_{03}(0)-\pi/2|,
$$
$$
\delta_{22}(\tau)=|\phi_{22}(\tau)-\phi_{22}(0)|,\quad
\delta_{23}(\tau)=|\phi_{23}(\tau)-\phi_{23}(0)|,\quad	
\delta_{33}(\tau)=|\phi_{33}(\tau)-\phi_{33}(0)|,\eqno(3.7)
$$
and $\delta_{ij}(\tau)=|\phi_{ij}(\tau)|$, for $15\ge i,j\ge 4$.
In this sub-section the initial conditions for the RDM were chosen to be,
$$
r_{ij}(0)=r_{ij}e^{i\pi/4},\quad r_{ji}(0)=r^*_{ij},\quad i<j,\quad 
r_{ii}(0)=r_{ii},\quad (i,j=0-3),\eqno(3.8)
$$
where $r_{ii}$ and $r_{ij}$ are taken from the RDM (3.3). To study the influence of non-resonant effects, we introduce the dependence of the qubit frequencies, $\omega_k$, on the parameter, $M$, according to the relation,
$$
\omega_k=\omega_0+M\cdot k,\quad (k=0,1,2,3).\eqno(3.9)
$$
Note, that in (3.1), $M=100$, and the frequencies of qubits were well-separated.
 In Fig.  4, we show the dependence of $\Delta_{ij}(\tau)$ on $M$ at the end of a $\pi$-pulse for the initial conditions (3.8), and for the following values of the parameters,
$$
J_{\alpha,\beta}=J=10,\quad\omega=(E_1-E_0)/\hbar=\omega_0+3J,\quad \Omega_k=\Omega=0.1.\eqno(3.10)
$$
The duration of the $\pi$-pulse is defined in (2.13). As one can see in Fig.  4, for $M> 40$, the values of $\Delta_{ij}(\tau)$ are small: $\Delta_{ij}(\tau)<0.01$. As numerical calculations show, for the chosen values of parameters (3.10), the critical value of $M$ is $M_{cr}\approx 30$. For $M<M_{cr}$, the value of $\Delta_{ij}(\tau)$ increases as $M$ decreases. In this case, the influence of non-resonant effects becomes significant, and the quantum dynamics of the system described by the Hamiltonian (2.3) does not correspond to the quantum CN operation defined in (2.10). In Fig. 5, the dependence $\delta_{ij}(\tau)$ on $M$ is shown for the same initial conditions and parameters used in Fig.  4. As one can see, the critical value $M_{cr}$ is of the same order, $M_{cr}\approx 30$. 

In the numerical calculations presented in Figs 6 and 7, the value of $M$ was fixed ($M=30$) and the value of the Ising constant, $J$, was varied. All other parameters and the initial conditions were chosen as in Figs. 4 and 5. In this case, the value, $M=30$ is close to the critical value, $M_{cr}$. As one can see from Figs. 6 and 7, for $J>1$ the quantum dynamics of the system corresponds to the dynamics of the desired quantum CN operation within an accuracy of about 1\%. For $J<J_{cr}\approx 0.3$, the influence of non-resonant effects becomes significant and one cannot use the system under consideration for quantum computation.\\ \ \\
{\bf 4. Conclusion}\\ \ \\
In this paper, we studied numerically the dynamics of a single-pulse quantum CN gate in an ensemble of four-spin molecules at room temperature. Our calculations confirm the equivalence between the action of a single-pulse quantum CN gate in an ensemble of four-spin molecules and in a pure two-spin quantum system studied earlier in \cite{berm1}, Chapter 25. Our results demonstrate the feasibility implementing of a single-pulse quantum CN gate at room temperature. 

We also have studied the influence  of non-resonant effects by varying the values of the separation between the resonant frequencies of spins, $\Delta\omega$, and the constant of the Ising interaction, $J$. We have found that non-resonant effects do not destroy a single-pulse quantum CN gate when the ratio of $\Delta\omega$ to the Rabi frequency, $\Omega$, is more than 300: $\Delta\omega/\Omega>300$, and for a wide range of the values of the Ising interaction constant, $J$: $3<J/\Omega<100$. Consequently, a single-pulse quantum CN gate can be experimentally realized if the transversal relaxation time, $T_2$, is bigger than 2$\pi/J$.

The results of this paper can be used for experimental implementation of a single-pulse quantum CN gate. In particular, a single-pulse quantum 
CN gate could be implemented in a liquid state NMR at room temperature and in solid-state systems in which the selective excitations of the type discussed in \cite{lloyd1,lloyd2,kane,cl} could be realized.\\ \ \\
{\bf Acknowledgments}\\ \ \\
G.V.L. and V.I.T. thank the Theoretical Division and the CNLS of the Los Alamos National Laboratory for their hospitality. This work  was supported by the Department of Energy under contract W-7405-ENG-36, and by the National Security Agency.
\newpage
\begin{center}
{\bf Figure Captions:} 
\end{center}
Figs. 1.  Dynamics of a quantum CN gate at room temperature for superpositional initial conditions (3.3). The matrix elements $r_{ii}$ for $i\ge 4$ are given in (2.18). The parameters are presented in (3.1). A $\pi$-pulse was applied with frequency  $\omega_0+3J$. Time evolution for the first four diagonal elements of the density matrix is shown. The left side shows the real part of the density matrix elements. The right side shows the imaginary part. \\ \ \\
Fig. 2. Time evolution of the non-diagonal matrix elements, $r_{01}$, $r_{02}$, 
 and $r_{03}$ for superpositional initial conditions. The parameters and initial conditions are the same as in Fig. 1.\\ \ \\
Fig. 3. Time-evolution of the non-diagonal matrix elements, $r_{12}$, $r_{13}$, and $r_{23}$ for superpositional initial conditions.  Parameters and initial conditions are the same as in Fig. 1.\\ \ \\
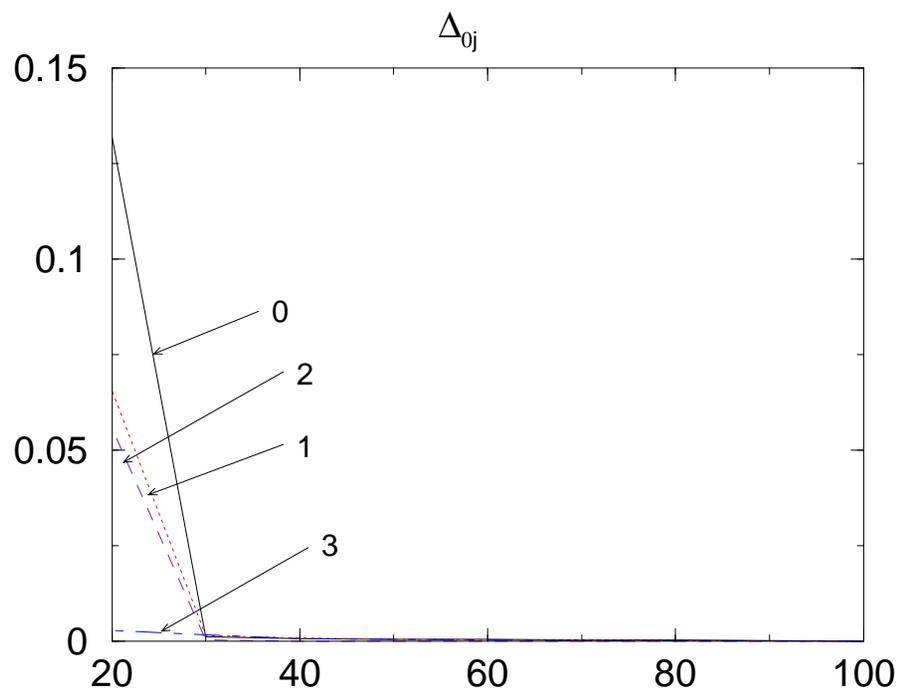
Fig.  4. Influence of non-resonant effects on dynamics of a quantum CN gate. Dependences of $\Delta_{0,j}$, $\Delta_{1,j}$,   $\Delta_{2,j}$, $\Delta_{11,7}$, and $\Delta_{3,3}$ on $M$, at the end of the $\pi$-pulse. The initial conditions are given in (3.8). Parameters are given in (3.9) and (3.10).
\\ \ \\
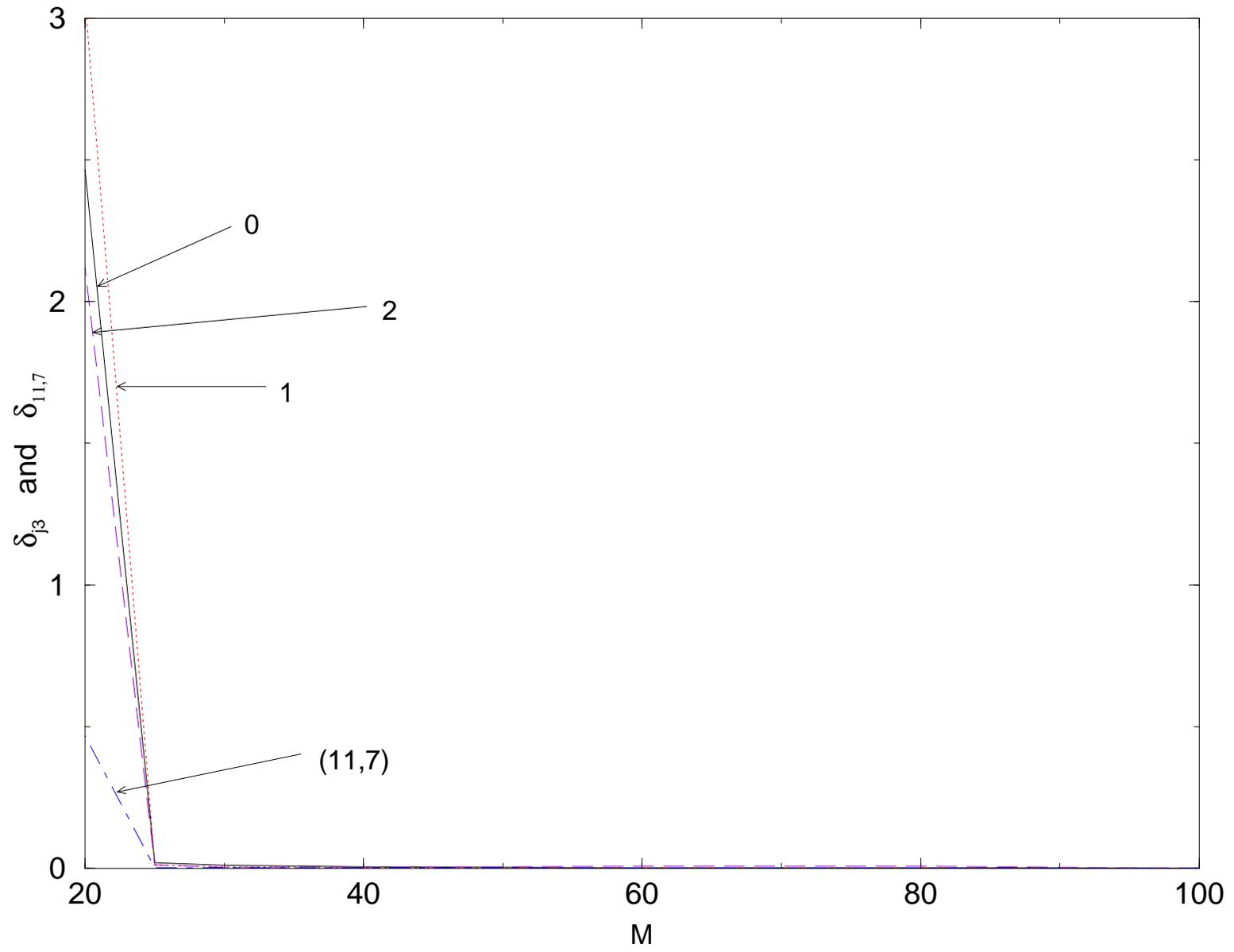
Fig.  5.  Influence of non-resonant effects on dynamics of a quantum CN gate. Dependences of $\delta_{j,3}$, and $\delta_{11,7}$ on $M$. Parameters and initial conditions are the same as in Fig.  4.\\ \ \\
Fig.  6. Influence of non-resonant effects on dynamics of a quantum CN gate. Dependences of $\Delta_{0,j}$, $\Delta_{1,j}$,  $\Delta_{2,j}$, $\Delta_{11,7}$, and $\Delta_{3,3}$ on $J$ (and $M=30$), at the end of the $\pi$-pulse. The initial conditions and the parameters are the same as in Fig.  4.
\\ \ \\
Fig.  7.  Influence of non-resonant effects on dynamics of a quantum CN gate. Dependences of $\delta_{j,3}$ on $J$ ($M=30$).
The initial conditions and the parameters are the same as in Fig.  4.\\ \ \\
\newpage

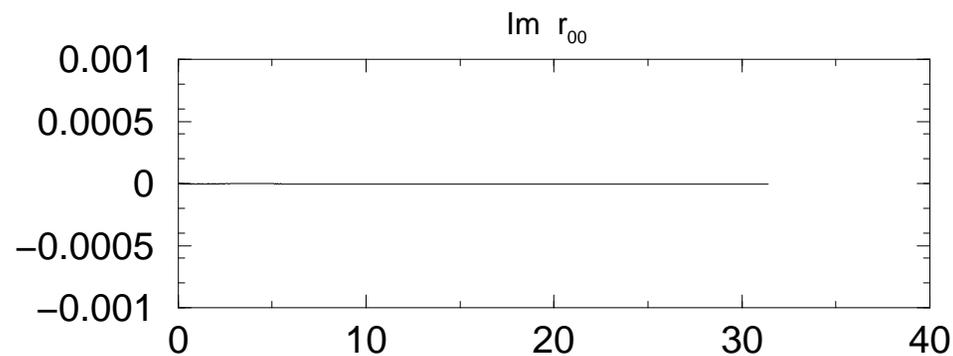
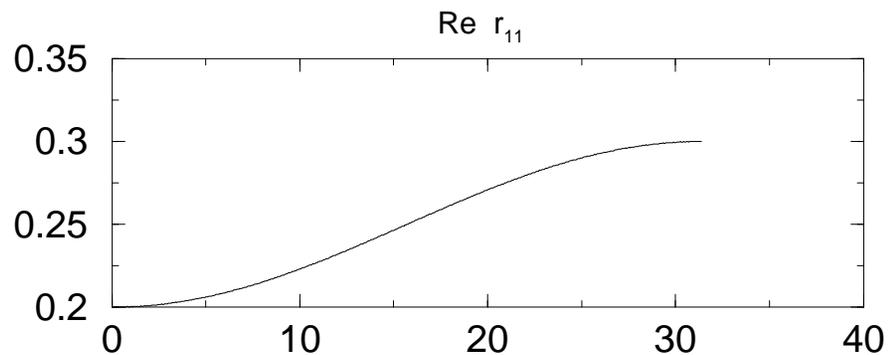
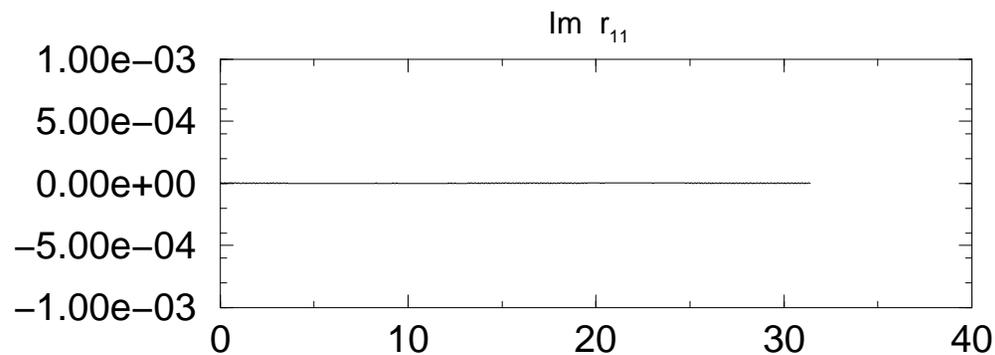
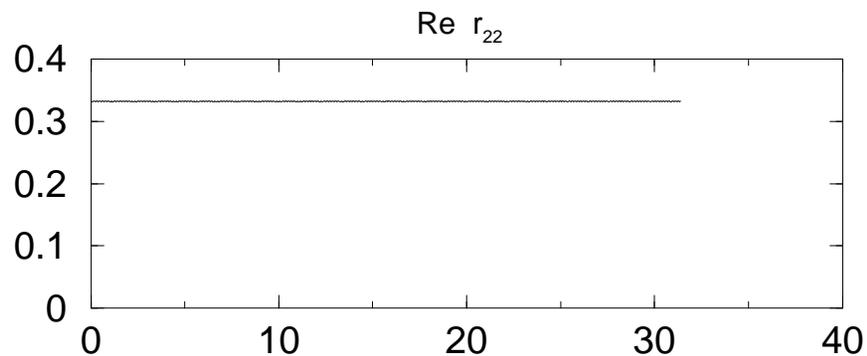
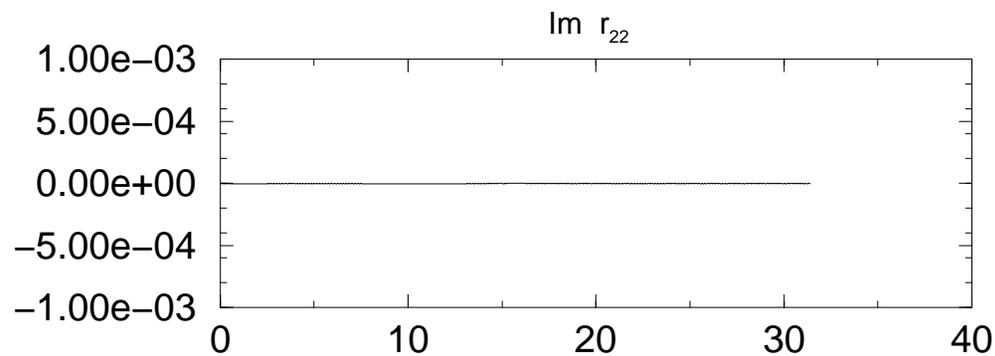
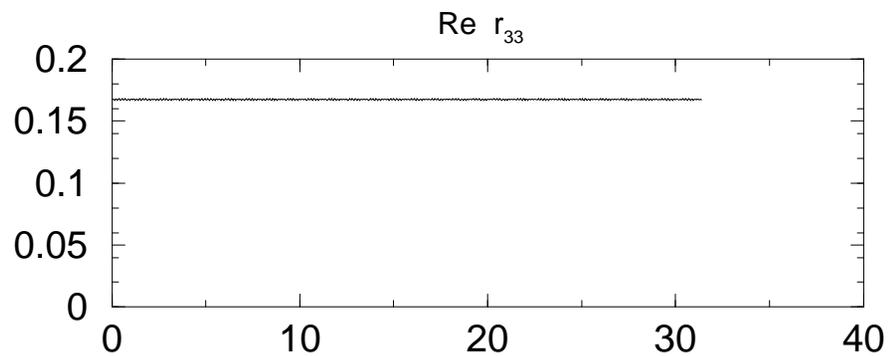
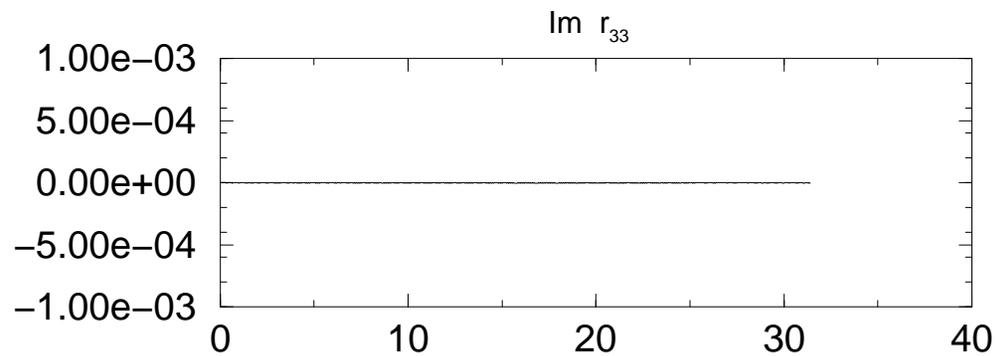
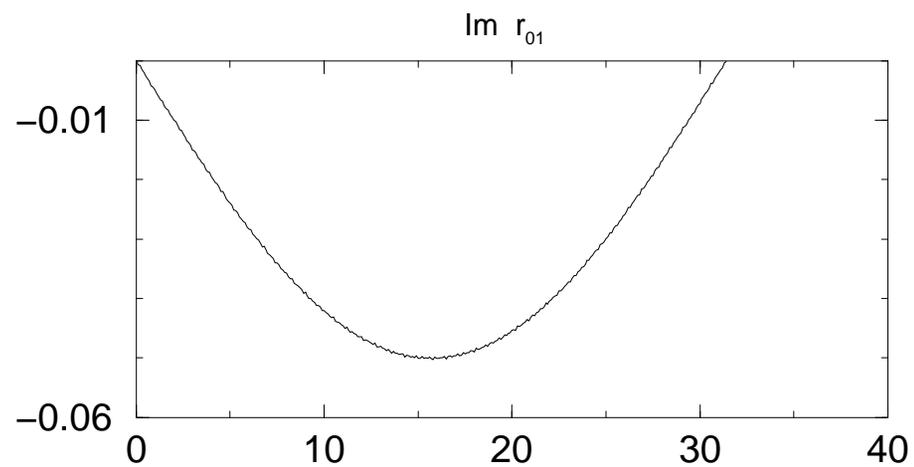
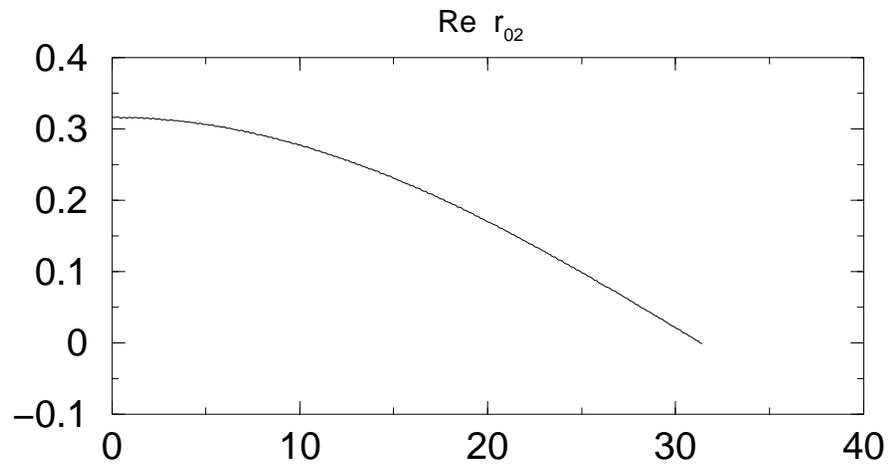
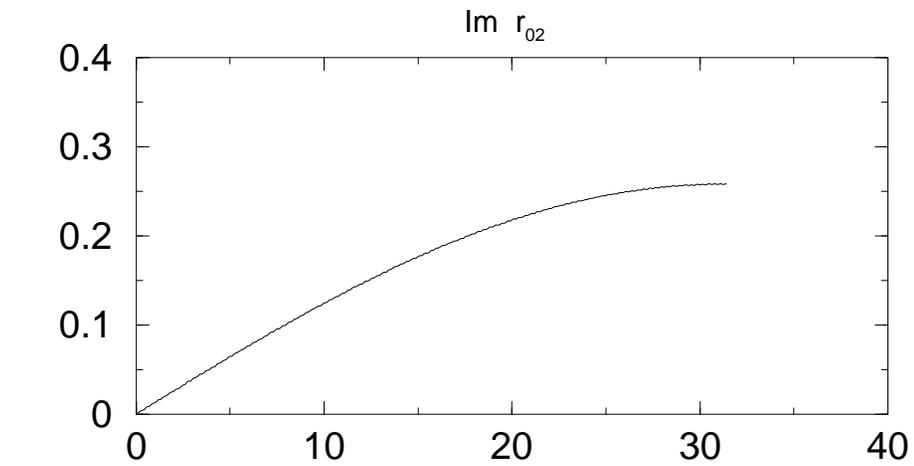
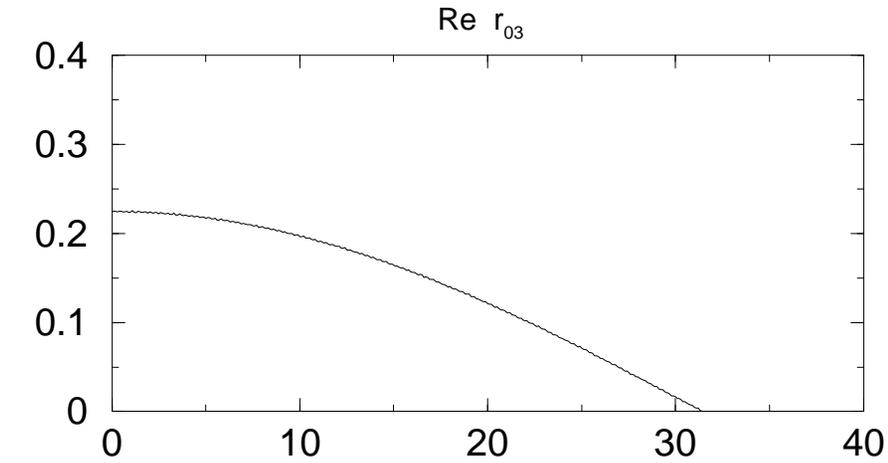
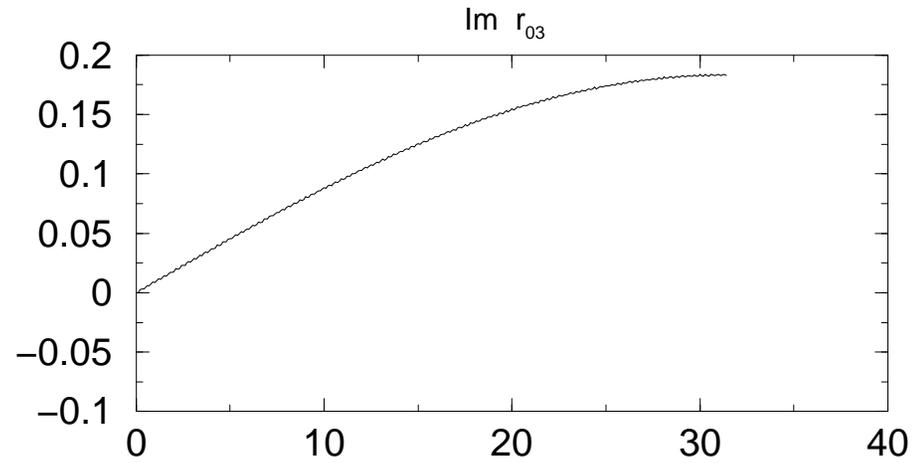
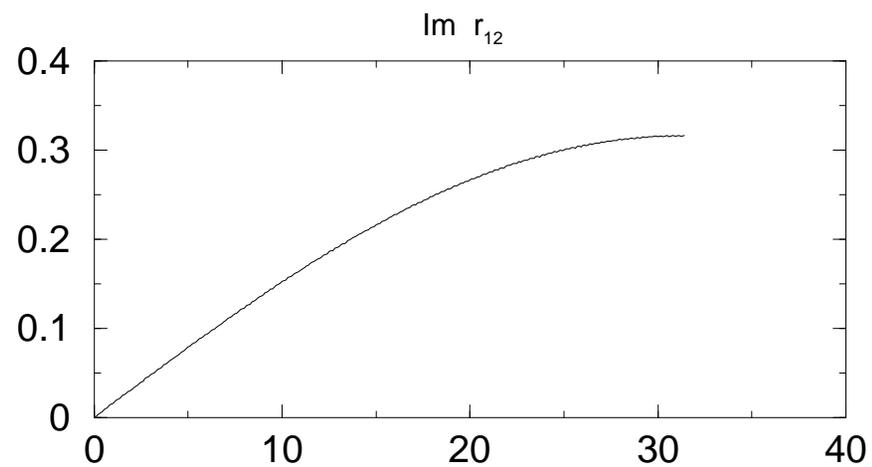
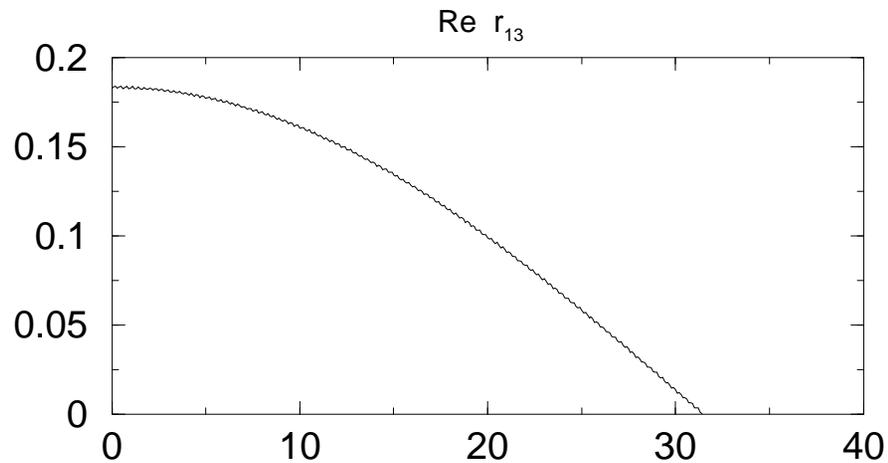
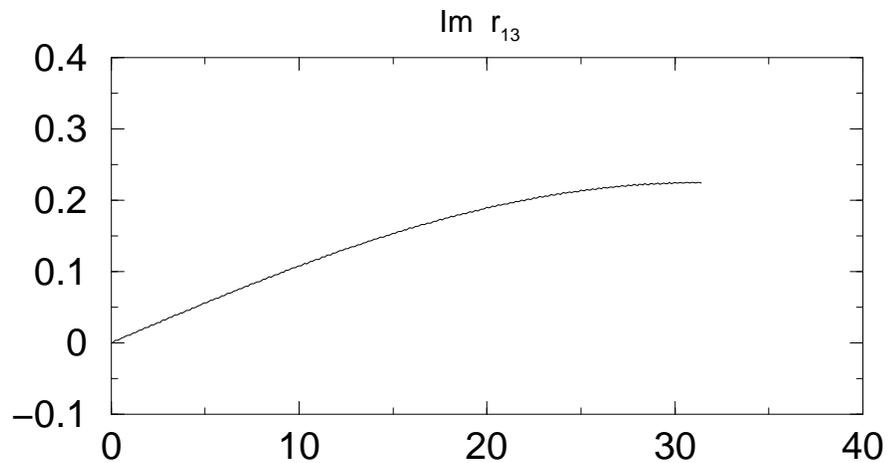
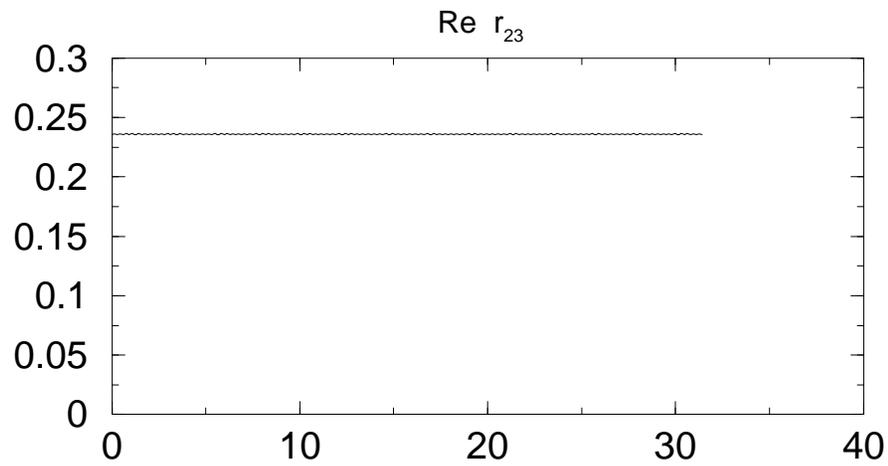
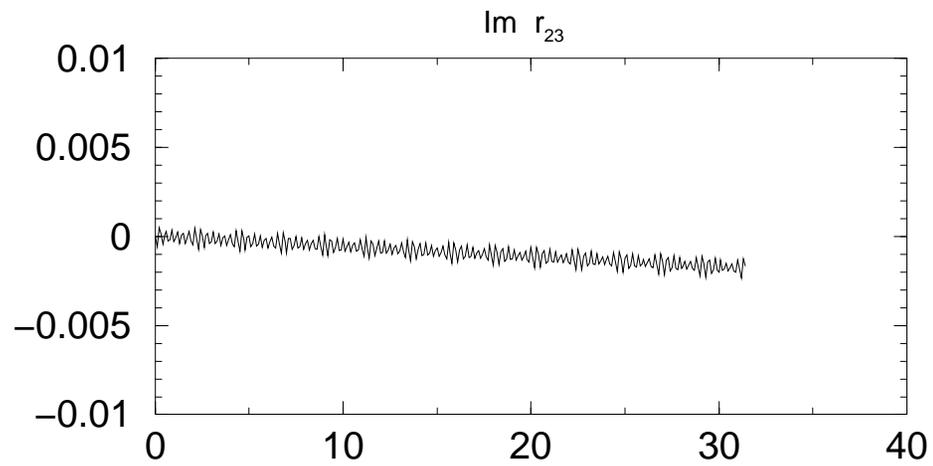
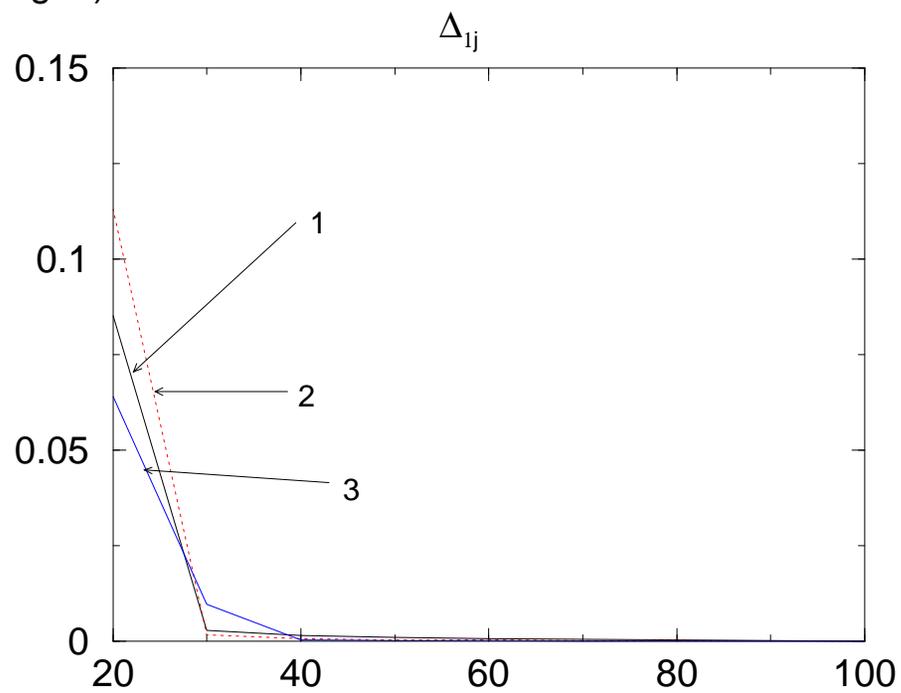
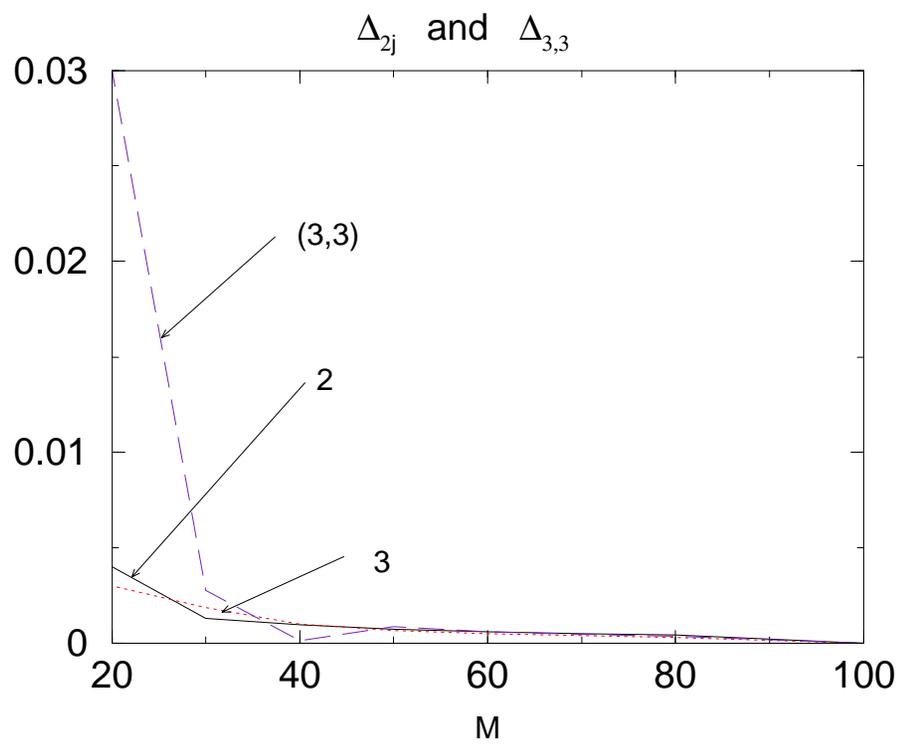
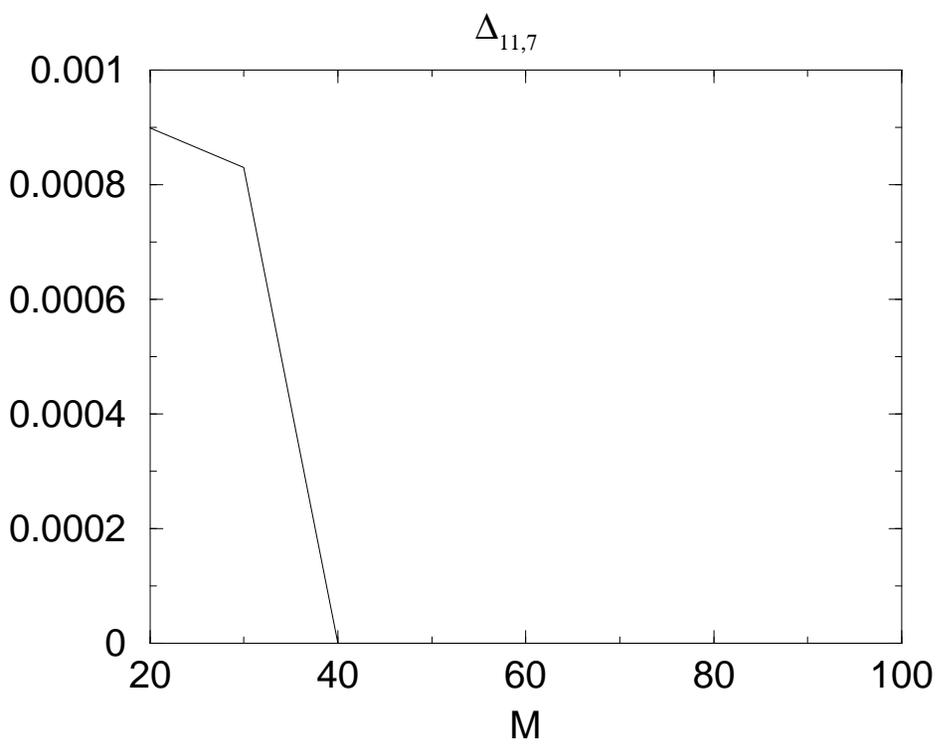
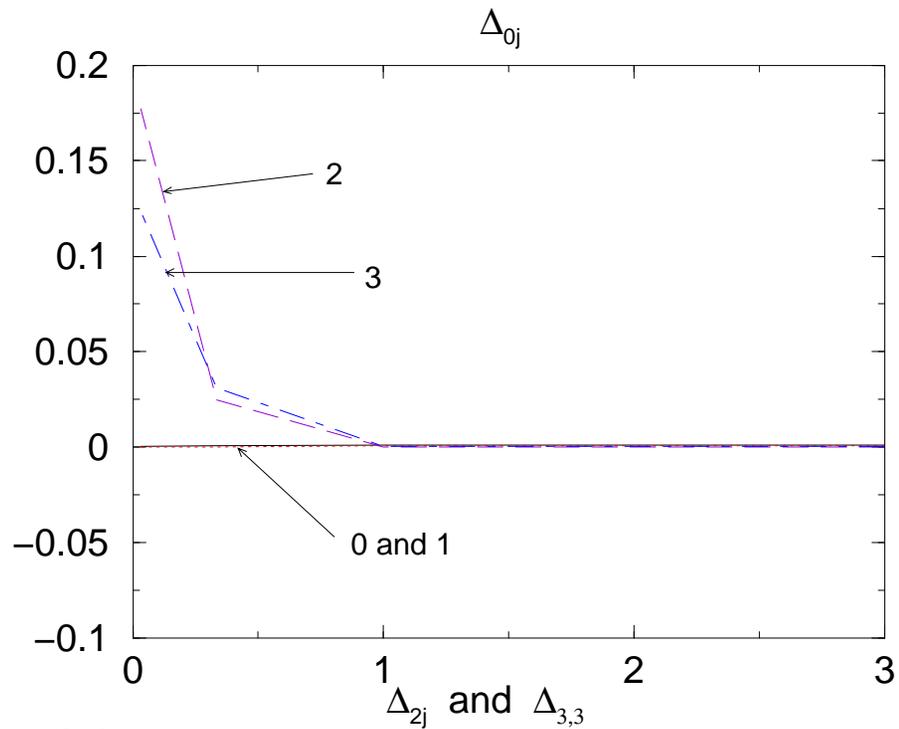
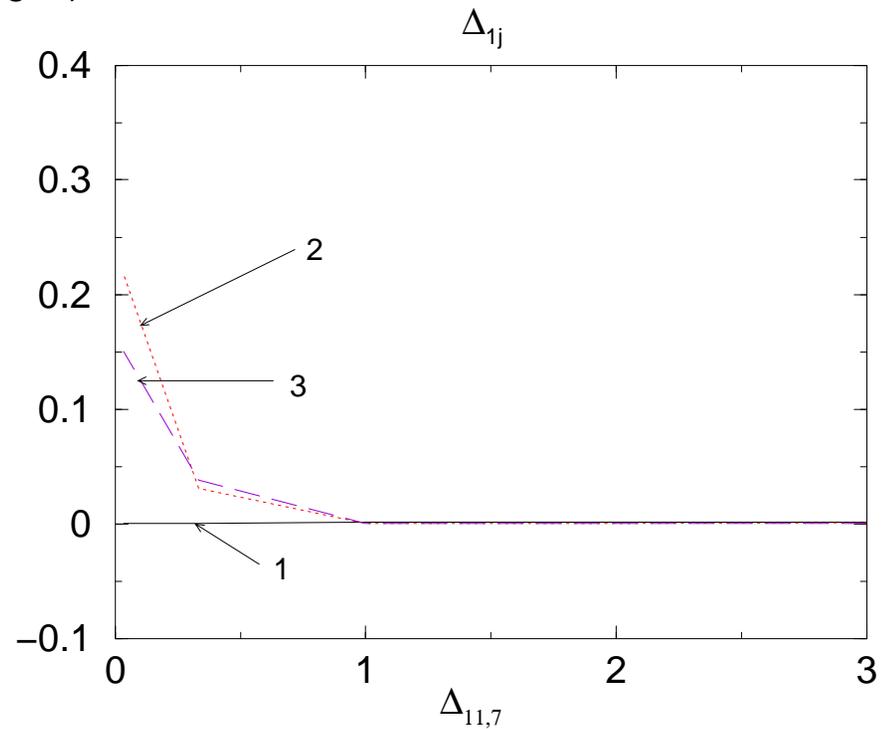
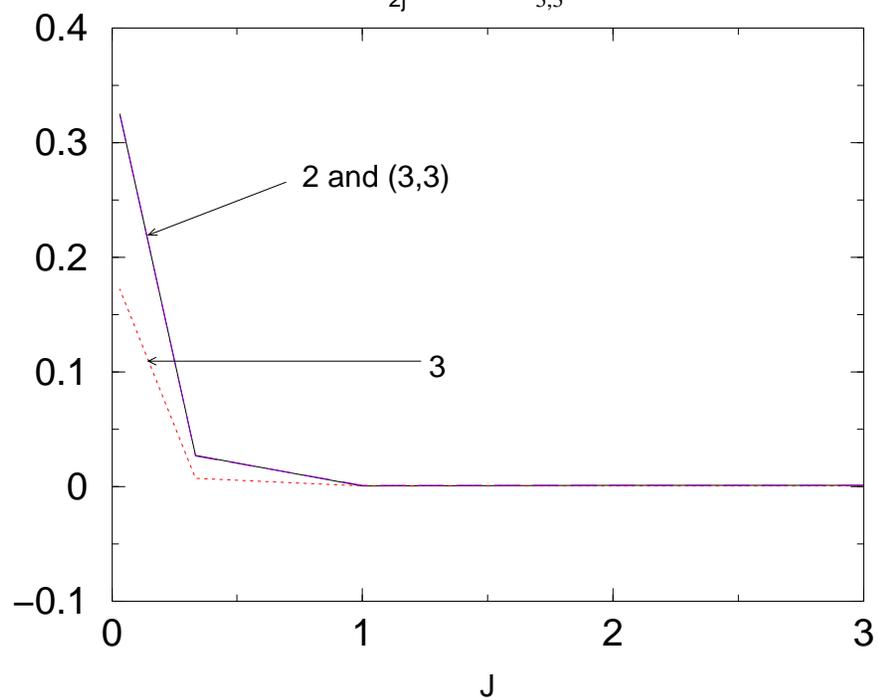
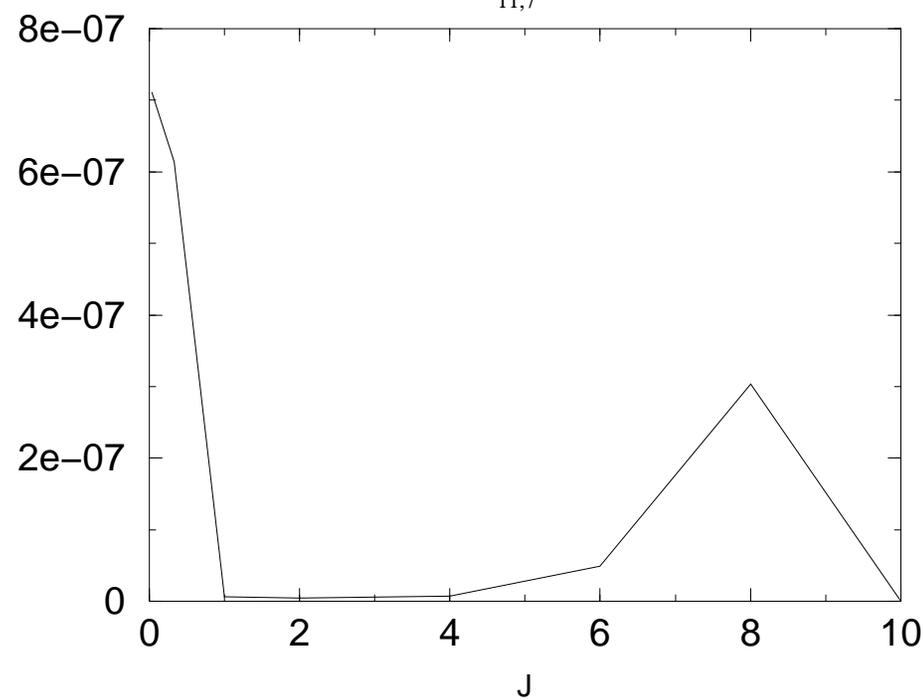
\end{document}